\documentclass[10pt]{article}
\usepackage{graphicx}
\usepackage[cp1251]{inputenc}
\usepackage[russian]{babel}

\begin{document}
\twocolumn [ \noindent {\footnotesize\it  ISSN 1063-7737,
 Astronomy Letters, 2006, Vol. 32, No. 9, pp. 608--621.
 \copyright Pleiades Publishing, Inc., 2006.

\noindent Original Russian Text \copyright V.V. Bobylev, M.Yu.
Khovritchev, 2006, published in Pis'ma v
Astronomicheski$\check{\imath}$ Zhurnal, 2006, Vol. 32, No. 9, pp.
676--890.}

\vskip -4mm

\begin{tabular}{llllllllllllllllllllllllllllllllllllllllllllllll}
 & & & & & & & & & & & & & & & & & & & & & & & & & & & & & & & & & & & & & & & \\
\hline \hline
\end{tabular}

\vskip 1.5cm

 \centerline {\Large\bf Kinematic Control of the Inertiality of the System of Tycho-2}
 \centerline {\Large\bf  and UCAC2 Stellar Proper Motions}
 \bigskip
 \centerline {\large\bf V. V. Bobylev
  and M. Yu. Khovritchev}
 \medskip
 \centerline {\it Pulkovo Astronomical Observatory, Russian Academy of Sciences,}
 \centerline {\it Pulkovskoe shosse 65, St.Petersburg, 196140 Russia}
 \centerline {\small Received December 15, 2005}

 \medskip
 \centerline {\small E-mail: {\tt vbobylev@gao.spb.ru}}

 \bigskip

{\noindent {\bf Abstract---}\small Based on the Ogorodnikov-Milne
model, we analyze the proper motions of Tycho-2 and UCAC2 stars.
We have established that the model component that describes the
rotation of all stars under consideration around the Galactic y
axis differs significantly from zero at various magnitudes. We
interpret this rotation found using the most distant stars as a
residual rotation of the ICRS/Tycho-2 system relative to the
inertial reference frame. For the most distant ($d\approx900$~pc)
Tycho-2 and UCAC2 stars, the mean rotation around the Galactic y
axis has been found to be $M_{13}=-0.37\pm0.04$~mas yr$^{-1}$. The
proper motions of UCAC2 stars with magnitudes in the range
$12-15^m$ are shown to be distorted appreciably by the magnitude
equation in $\mu_\alpha\cos\delta$, which has the strongest effect
for northern-sky stars with a coefficient of $-0.60\pm0.05$~mas
yr$^{-1}$ mag$^{-1}$. We have detected no significant effect of
the magnitude equation in the proper motions of UCAC2 stars
brighter than $\approx11^m$.
}
\bigskip


{\bf DOI}: 10.1134/S1063773706090064

\vskip 1cm

]

\section*{INTRODUCTION}

The current standard International Celestial Reference System
(ICRS) is realized by a catalog of radio positions for 212 compact
extragalactic sources uniformly distributed over the entire sky
and observed by very long baseline interferometry (Ma et al.
1998). In the optical range, the first ICRS realization was the
Hipparcos catalog (ESA 1997). Application of various methods of
analysis reveals a small residual rotation of ICRS/Hipparcos
relative to the inertial reference system with
$\approx-0.4\pm0.1$~mas yr$^{-1}$ (Bobylev 2004a, 2004b).

The method considered here is based on the study of two solid-body
rotation tensor components that describe the rotation around the y
and x axes in the Galactic coordinate system. Application of this
method to the proper motions of TRC stars (H{\o}g et al. 1998)
showed that the determination of the rotation around the y axis is
noticeably affected by the actual rotation of the Local system's
stars (Bobylev 2004a). In this paper, we consider the Tycho-2
catalog (H{\o}g et al. 2000), which does not differ systematically
from the TRC catalog, but contains twice as many stars.  We expect
a confirmation of the results obtained using TRC stars.

Invoking fairly distant stars that are free from the effects of
both streams close to the Sun and the Local system of stars as a
whole is topical for a reliable application of the method. The
UCAC2 catalog (Zacharias et al. 2004), which contains positions
and proper motions for some 48 million stars, is of great interest
in this connection. This catalog extends the ICRS/Tycho-2 system
to the 17~th magnitude.

The goal of this paper is to study the kinematic parameters of a
large number of stars as a function of their distance. To estimate
the distances, we use the method of comparing the statistical
parallaxes with the solar velocity (Olling and Dehnen 2003); the
latter is currently known well (Dehnen and Binney 1998).

\section{THE MODEL}

In this paper, we use a rectangular Galactic coordinate system
with the axes directed away from the observer toward the Galactic
center $(l=0^\circ, b=0^\circ$, the $x$ axis or axis 1), along the
Galactic rotation $(l=90^\circ, b=0^\circ$, the $y$ axis or axis
2), and toward the North Galactic pole $(b=90^\circ$, the $z$ axis
or axis 3). In the Ogorodnikov-Milne model, we use the notation
introduced by Clube (1972, 1973) and employed by du Mont (1977,
1978). When only the stellar proper motions are used, one of the
diagonal terms of the local deformation tensor is known
(Ogorodnikov 1965) to remain indeterminate. Therefore, we
determine differences of the form
$(M_{\scriptscriptstyle11}^{\scriptscriptstyle+}-
 M_{\scriptscriptstyle22}^{\scriptscriptstyle+})$ and
$(M_{\scriptscriptstyle33}^{\scriptscriptstyle+}-
 M_{\scriptscriptstyle22}^{\scriptscriptstyle+})$.

The conditional equations can be written as
$$\displaylines{\hfill
\mu_{l}\cos b=
       (1/r)(X_{\odot}\sin l-Y_{\odot}\cos l)-
\hfill\llap(1) \cr\hfill
   -M_{\scriptscriptstyle32}^{\scriptscriptstyle-}\cos l\sin b
   -M_{\scriptscriptstyle13}^{\scriptscriptstyle-}\sin l\sin b
   +M_{\scriptscriptstyle21}^{\scriptscriptstyle-}\cos b+
\hfill\cr\hfill
   +M_{\scriptscriptstyle12}^{\scriptscriptstyle+}\cos 2l\cos b
   -M_{\scriptscriptstyle13}^{\scriptscriptstyle+}\sin l\sin b+
\hfill\cr\hfill
   +M_{\scriptscriptstyle23}^{\scriptscriptstyle+}\cos l\sin b-
\hfill\cr\hfill
  -0.5(M_{\scriptscriptstyle11}^{\scriptscriptstyle+}
  -M_{\scriptscriptstyle22}^{\scriptscriptstyle+})\sin 2l\cos b,
\hfill \cr\hfill
\mu_b=
    (1/r)(X_{\odot}\cos l\sin b+Y_{\odot}\sin l\sin b-
\hfill\llap(2) \cr\hfill
    -Z_{\odot}\cos b)
   +M_{\scriptscriptstyle32}^{\scriptscriptstyle-}\sin l
   -M_{\scriptscriptstyle13}^{\scriptscriptstyle-}\cos l-
\hfill\cr\hfill
-0.5M_{\scriptscriptstyle12}^{\scriptscriptstyle+}\sin 2l\sin 2b
   +M_{\scriptscriptstyle13}^{\scriptscriptstyle+}\cos l\cos 2b+
\hfill\cr\hfill
   +M_{\scriptscriptstyle23}^{\scriptscriptstyle+}\sin l\cos 2b-
\hfill\cr\hfill
-0.5(M_{\scriptscriptstyle11}^{\scriptscriptstyle+}
    -M_{\scriptscriptstyle22}^{\scriptscriptstyle+})\cos^2 l\sin 2b+
\hfill\cr\hfill
+0.5(M_{\scriptscriptstyle33}^{\scriptscriptstyle+}
    -M_{\scriptscriptstyle22}^{\scriptscriptstyle+})\sin 2b.
\hfill }
$$
where $X_\odot,Y_\odot,Z_\odot$ are the peculiar solar velocity
components and $M_{\scriptscriptstyle12}^{\scriptscriptstyle-},
M_{\scriptscriptstyle13}^{\scriptscriptstyle-},
M_{\scriptscriptstyle23}^{\scriptscriptstyle-}$ are the components
of the vector of solid-body rotation of a small solar neighborhood
around the corresponding axes. In accordance with the adopted
rectangular coordinate system, the following rotations are
positive: from axis 1 to axis 2, from axis 2 to axis 3, and from
axis 3 to axis 1. The quantity
$M_{\scriptscriptstyle21}^{\scriptscriptstyle-}$ (mas yr$^{-1}$)
is related to the Oort constant $B$ (km s$^{-1}$ kpc$^{-1}$) via
the proportionality coefficient 4.74. Each of the quantities
 $M_{\scriptscriptstyle12}^{\scriptscriptstyle+}$,
 $M_{\scriptscriptstyle13}^{\scriptscriptstyle+}$ and
 $M_{\scriptscriptstyle23}^{\scriptscriptstyle+}$ describes the
deformation in the corresponding plane. The quantity
 $M_{\scriptscriptstyle12}^{\scriptscriptstyle+}$ (mas yr$^{-1}$)
is related to the Oort constant A (km s$^{-1}$ kpc$^{-1}$) via the
proportionality coefficient 4.74.
The diagonal components of the
local deformation tensor
 $M_{\scriptscriptstyle11}^{\scriptscriptstyle+}$,
 $M_{\scriptscriptstyle22}^{\scriptscriptstyle+}$ and
 $M_{\scriptscriptstyle33}^{\scriptscriptstyle+}$ describe the
overall contraction or expansion of the entire stellar system. The
system of conditional equations (1) and (2) includes eleven
sought-for unknowns to be determined by the least squares method.
As can be seen from Eq. (1), the two pairs of unknowns
 $M_{\scriptscriptstyle13}^{\scriptscriptstyle-}$ and
 $M_{\scriptscriptstyle13}^{\scriptscriptstyle+}$ as well as
 $M_{\scriptscriptstyle32}^{\scriptscriptstyle-}$ and
 $M_{\scriptscriptstyle23}^{\scriptscriptstyle+}$
have the same coefficients, $\sin l\sin b$ and $\cos l\sin b$,
respectively. As a result, the variables are ill-separated. In the
UCAC2 section, we analyze the results of both the  simultaneous
solution of Eqs. (1) and (2) and the separate solution of only Eq.
(2). The quantity $1/r$ is a parallactic factor that is taken to
be unity when solving Eqs. (1) and (2). In this case, the stars
are referred to a unit sphere. In this approach, all of the
parameters being determined are proportional to the heliocentric
distance of the stellar centroid under consideration and are
expressed in the same units as the stellar proper motion
components, i.e., in mas yr$^{-1}$. Using this approach, we can
completely eliminate the effect of distance errors in the data
being analyzed. Indeed, when using the method with known distances
to stars, we must multiply the left-hand and right-hand sides of
Eqs. (1) and (2) by $4.74r$ and $r$, respectively; the unknowns
being determined will then be distorted by errors in the distances
to stars. At present, reliable distances to individual stars
 (with errors $<10\%$)
allow us to analyze a solar neighborhood $\sim$100~pc in radius,
which is not enough for our purposes.

Let us also consider a Galactocentric cylindrical $(R,\theta,z)$
coordinate system specified as follows: the $z$ axis is directed
toward the North Galactic Pole from the Galactic center; the
azimuthal angle $\theta$ is measured from the $x$ axis to the $y$
axis counterclockwise, and $R$ is the Galactocentric distance of a
star. In the cylindrical coordinate system, the local deformation
and local rotation tensor components are
$$\displaylines{\hfill
M_{\scriptscriptstyle11}^{\scriptscriptstyle+}=
  {\strut\displaystyle \partial\displaystyle V_R\over\displaystyle\partial\displaystyle R},
\hfill\cr\hfill
 M_{\scriptscriptstyle22}^{\scriptscriptstyle+}=
 {\strut\displaystyle 1\over\displaystyle R}{\strut\displaystyle\partial\displaystyle V_\theta\over\displaystyle\partial\displaystyle  \theta}+{\strut\displaystyle  V_R\over\displaystyle R}, \hfill\cr\hfill
 M_{\scriptscriptstyle33}^{\scriptscriptstyle+}=
  {\strut\displaystyle \partial\displaystyle  V_z\over\displaystyle\partial\displaystyle  z},
\hfill\cr\hfill
 M_{\scriptscriptstyle12}^{\scriptscriptstyle+}=
 0.5\biggl(
 {\strut\displaystyle 1\over\displaystyle  R}{\strut\displaystyle\partial\displaystyle V_R\over\displaystyle\partial\displaystyle  \theta}
 -{\strut\displaystyle  V_\theta\over\displaystyle R}
 +
 {\strut\displaystyle \partial\displaystyle  V_\theta\over\displaystyle\partial\displaystyle  R}
 \biggr),
\hfill \cr\hfill
 M_{\scriptscriptstyle13}^{\scriptscriptstyle+}=
 0.5\biggl(
 {\strut\displaystyle \partial\displaystyle  V_R\over\displaystyle\partial\displaystyle z}
 +
 {\strut\displaystyle \partial\displaystyle  V_z\over\displaystyle\partial\displaystyle  R}
 \biggr),
\hfill\llap(3)\cr\hfill
 M_{\scriptscriptstyle23}^{\scriptscriptstyle+}=
 0.5\biggl(
 {\strut\displaystyle \partial\displaystyle  V_\theta\over\displaystyle\partial\displaystyle z}
 +
 {\strut\displaystyle 1\over\displaystyle R}{\strut \displaystyle \partial\displaystyle V_z\over\displaystyle\partial\displaystyle \theta}
 \biggr),
\hfill\cr\hfill
 M_{\scriptscriptstyle21}^{\scriptscriptstyle-}=
 0.5\biggl(
 {\strut\displaystyle \partial\displaystyle  V_\theta\over\displaystyle\partial\displaystyle  R}
 -
 {\strut\displaystyle 1\over\displaystyle  R}{\strut\displaystyle\partial\displaystyle V_R\over\displaystyle\partial\displaystyle  \theta}
 +{\strut\displaystyle  V_\theta\over\displaystyle R}
 \biggr),
\hfill \cr\hfill
 M_{\scriptscriptstyle32}^{\scriptscriptstyle-}=
 0.5\biggl(
{\strut\displaystyle 1\over\displaystyle R}{\strut \displaystyle
\partial\displaystyle V_z\over\displaystyle\partial\displaystyle
\theta}
 -
 {\strut\displaystyle \partial\displaystyle V_\theta\over\displaystyle\partial\displaystyle z}
 \biggr),
\hfill\cr\hfill
 M_{\scriptscriptstyle13}^{\scriptscriptstyle-}=
 0.5\biggl({\strut\displaystyle \partial\displaystyle  V_R\over\displaystyle\partial\displaystyle z}-
     {\strut\displaystyle \partial\displaystyle  V_z\over\displaystyle\partial\displaystyle  R}
     \biggr), \hfill
 }
$$
provided that the derivatives are taken at point
$(R_\circ,\theta_\circ,z_\circ)=(R_\circ,0^\circ,0)$.

\section{TYCHO-2}

The results of solving the system of equations (1) and (2) using
Tycho-2 stars of mixed spectral composition as a function of
magnitude are presented in Fig. 1. We used almost all stars of the
catalog with the following constraint imposed on the absolute
value of a star's tangential velocity: $|\mu_t|=\sqrt{{\mu_\alpha
\cos\delta}^2+\mu_\delta^2}<300$~mas yr$^{-1}$. The stars were
divided into magnitude ranges in such a way that each of them
contained approximately the same number of stars ($\approx$250
000). In each magnitude range, the random errors of all sought-for
parameters are $\approx$0.05 mas yr$^{-1}$; the errors in
 $(M_{\scriptscriptstyle11}^{\scriptscriptstyle+}-
   M_{\scriptscriptstyle22}^{\scriptscriptstyle+})$ and
 $(M_{\scriptscriptstyle33}^{\scriptscriptstyle+}-
   M_{\scriptscriptstyle22}^{\scriptscriptstyle+})$
are twice as large. As can be seen from Fig. 1, the parameters
that describe the deformation in the $yz$ plane and the rotation
around the $x$ axis, i.e.,
 $M_{\scriptscriptstyle23}^{\scriptscriptstyle+}$ and
 $M_{\scriptscriptstyle32}^{\scriptscriptstyle-}$,
are almost equal to zero. The parameters that describe the
deformations in the $xy$ and $yz$ planes and the rotations around
the $z$ and $y$ axes differ significantly from zero. The parameter
$M_{\scriptscriptstyle13}^{\scriptscriptstyle-}$ is nonzero up to
the faintest Tycho-2 stars. The magnitude dependence of
$M_{\scriptscriptstyle13}^{\scriptscriptstyle-}$ agrees well with
that found using the TRC catalog, where the mean is
 $M_{\scriptscriptstyle13}^{\scriptscriptstyle-}=-0.86\pm0.11$~mas yr$^{-1}$
(Bobylev 2004a). For Tycho-2 stars fainter than $8^m.5$, the mean
value (the mean often points) is
 $M_{\scriptscriptstyle13}^{\scriptscriptstyle-}=-0.88\pm0.10$~mas yr$^{-1}$.
This is  because the fraction of stars belonging to the Local
system or local nearby streams is large even among the faintest
$(\approx13^m)$ Tycho-2 stars.

We obtained the following Galactic rotation parameters from stars
fainter than $8^m.5$:
 $M_{\scriptscriptstyle12}^{\scriptscriptstyle+}= 2.93\pm0.21$~mas yr$^{-1}$ and
 $M_{\scriptscriptstyle21}^{\scriptscriptstyle-}=-2.35\pm0.10$~mas yr$^{-1}$.
In this case, the Oort constants are
 $A=4.74 M_{\scriptscriptstyle12}^{\scriptscriptstyle+}= 13.88\pm0.98$ km s$^{-1}$ kpc$^{-1}$ and
 $B=4.74 M_{\scriptscriptstyle21}^{\scriptscriptstyle-}=-11.13\pm0.47$ km s$^{-1}$ kpc$^{-1}$.
These Oort constants serve as an indicator of the consistency of
our approach (constraints, sky coverage). On the whole, they are
in agreement, for example, with the results of analyzing the
proper motions of ACT/Tycho-2 stars (Olling and Dehnen 2003),
where only equatorial stars $(|b|<5.73^\circ)$ were used.

We use the proper motions of Tycho-2 stars to check whether
nonuniformities in the latitudinal distribution of star s affect
the determination of the Ogorodnikov-Milne model parameters. For
this purpose, we averaged the stellar proper motions in fields of
equal area that covered the entire sky. In contrast to 48 fields
used in the well-known Charlier method (Ogorodnikov 1965), we
divided the sky into 432 fields. The essence of the method is
that, despite the difference in the number of stars, a unit weight
is assigned to each field when solving the system of conditional
equations (1) and (2). As a result, the random errors of the main
sought-for parameters in each magnitude range were
 $\approx0.15$~mas yr$^{-1}$; no marked differences from the previous approach were found.
The parameter
 $M_{\scriptscriptstyle21}^{\scriptscriptstyle-}$, for
which the values calculated by this method are indicated in Fig. 1
by the solid line, constitutes an exception. As can be seen from
the figure, all of the points found have approximately the same
displacement of
 $\approx0.2$~mas yr$^{-1}$
along the coordinate axis. The mean Oort constant
 $B$ is $-12.13\pm0.63$ km s$^{-1}$ kpc$^{-1}$.
Application of the method with fields showed that the correlation
coefficients between all of the unknowns being determined do not
exceed 0.1, except the correlations between two unknowns,
 $(M_{\scriptscriptstyle11}^{\scriptscriptstyle+}-
 M_{\scriptscriptstyle22}^{\scriptscriptstyle+})$ and
 $(M_{\scriptscriptstyle33}^{\scriptscriptstyle+}-
 M_{\scriptscriptstyle22}^{\scriptscriptstyle+})$,
for which the correlation coefficient is 0.5. Below, we use the
method with individual stars and will return to our comparison of
the results obtained by the two methods.

\section{UCAC2}
\subsection{\it The First Approximation}

The UCAC2 astrometric catalog is currently the only mass catalog
that contains highly accurate proper motions of faint stars
(fainter than $12^m.5$) in much of the celestial sphere. There is
full sky coverage in the declination zone $-90^\circ <\delta
<40^\circ$ and partial sky coverage up to $\delta=52^\circ$. The
accuracy of the proper motions of faint UCAC2 stars given by the
authors of the catalog lies within the range 4 to 7 mas yr$^{-1}$.
The catalog extends the ICRS/Tycho-2 system to faint stars.

Since the catalog contains no observations at declinations
 $\delta>+60^\circ$, we do not consider the southernmost stars at declinations
 $\delta<-60^\circ$ in order that the sky coverage be symmetric.

The results of solving the system of equations (1) and (2) using
UCAC2 stars of mixed spectral composition are presented in Fig. 2
as a function of magnitude. We used a constraint on the absolute
of the tangential stellar velocity, $|\mu_t| < 300$~mas yr$^{-1}$.
There were $\sim$250 000 stars in each magnitude range. In the
range $12^m.05-16^m.01$, the stars were taken selectively. In the
ranges of magnitudes brighter than $11^m.17$ and fainter than
$16^m.34$, we used all stars from the catalog. Thus, we consider a
total of $\sim$4.4 million UCAC2 stars. As can be seen from Fig.
2, the parameters that describe the deformations in the $xy$,
$yz$, $zx$ planes and the rotation around the $z$ and $y$ axes
differ significantly from zero. The parameter
$M_{\scriptscriptstyle13}^{\scriptscriptstyle-}$ is nonzero for
the faintest UCAC2 stars.

As can be seen from Fig. 2, there is a noticeable jump in the
magnitude dependence of
 $M_{\scriptscriptstyle13}^{\scriptscriptstyle-}$ for stars
fainter than $13^m$, which is inconsistent with the assumption
that this  parameter monotonically tends (in absolute value) to a
minimum. This jump can be explained by the presence of a magnitude
equation in the proper motions of UCAC2 stars. Analysis of Eqs.
(1) and (2) in the equatorial coordinate system (du Mont 1977,
1978) leads us to conclude that the rotation around the Galactic y
axis can be affected by a systematic error only in
$\mu_\alpha\cos\delta$. Our simulation shows that the coefficient
of the magnitude equation for the entire catalog is 0.3$-$0.4~mas
yr$^{-1}$ mag$^{-1}$ for stars fainter than $13^m$. Eliminating
this magnitude equation yields
 $M_{\scriptscriptstyle13}^{\scriptscriptstyle-}\approx-1$~mas
yr$^{-1}$ for $13^m$ stars.

On the other hand, we can study the magnitude equation of the
UCAC2 catalog by an independent method, by comparison with the
original Pulkovo data. As Bobylev et al. (2004) showed, the
Pulkovo stellar proper motions do not have a noticeable magnitude
equation.

\subsection{\it The Magnitude Equation in UCAC2}

To find the magnitude equation in the proper motions of UCAC2
stars, we used the new proper motions of faint stars calculated
using the stellar coordinates from the Pul-3 catalog (Khrutskaya
et al. 2004) and currently available astrometric catalogs: M2000
(Rapaport et al. 2001), CMC13 (2003) and UCAC2.

The Pul-3 astrometric catalog was compiled at the Pulkovo
Observatory by measuring photographic plates taken with a normal
astrograph as part of A. N. Deutsch's plan to determine the
absolute proper motions of stars in fields with galaxies. The
catalog contains equatorial coordinates for more than 50 000 faint
stars to $16^m.5$ in 146 fields north of $\delta=-5^\circ$ for the
mean epoch of the Pulkovo observations (1963.25) in the
ICRS/Tycho-2 system. The positional accuracy of faint stars in the
catalog is 80 mas.

The M2000 (2.3 million stars) and CMC13 (36.2 million stars)
catalogs were compiled using CCD observations with the Bordeaux
and La Palma automated meridian circles. The two catalogs realize
the ICRS/Tycho-2 system in the declination zones
 $11^\circ-18^\circ$ and $-3^\circ-30^\circ$,
respectively, and contain stars to $16^m-17^m$. The positional
accuracy of faint stars in these catalogs is, on average, 40$-$60
mas at epoch 2000.

Mutual identifications of stars in these catalogs made it possible
to compile a new version of the catalog of proper motions for 34
000 stars. We designate the new version of the catalog as PUL3SE.
Owing to the high positional accuracy of the stars in the original
catalogs and the large epoch difference between Pul-3 and the
catalogs for epoch 2000, the estimated internal accuracies of the
proper motions ($\epsilon_{\mu_\alpha\cos\delta}$ и
 $\epsilon_{\mu_\delta}$) lie within the range 2 to 4 mas yr$^{-1}$.

For each star included in both catalogs, we formed the
PUL3SE-UCAC2 differences between the proper motions in both
coordinates. The mean differences for all data are 1 and
$-0.5$~mas yr$^{-1}$ for the proper motions in right ascension and
declination, respectively. The errors of a single difference were
found to be
 $\sigma_{\mu_\alpha\cos\delta}=6.3$~mas yr$^{-1}$
 and
 $\sigma_{\mu_\delta}=5.7$~mas yr$^{-1}$.
For the groups of stars formed as a function of magnitude U2Rmag
at $0^m.25$ steps, the ratios
 $(\epsilon_{\mu_\alpha\cos\delta}^2+
\epsilon_{\mu_\alpha\cos\delta\,ucac2}^2)/\sigma_{\mu_\alpha\cos\delta}^2$
 and
$(\epsilon_{\mu_\delta}^2+\epsilon_{\mu_\delta\,ucac2}^2)/\sigma_{\mu_\delta}^2$
are close to unity. This confirms that the formal estimates of the
accuracies are realistic for both the new proper motions and the
UCAC2 proper motions presented in this catalog.

The dependences of $\Delta\mu$ on magnitude U2Rmag are presented
in Fig.~3. Analysis of these dependences reveals no magnitude
equation in the proper motions of faint stars in both coordinates
for relatively bright stars (brighter than $12^m.5$). This may be
interpreted as a result of the fact that all of the catalogs used
to calculate the proper motions realize the ICRS/Tycho-2 system.
The Tycho-2 catalog is known to be complete to 11m and to contain
a considerable number of star s to $12^m.5$. For fainter stars,
systematic magnitude dependent errors in the coordinates show up
in different catalogs.

On the whole, we may conclude that the magnitude equation for the
proper motions of faint stars in declination is indistinct (about
$-0.5$~mas yr$^{-1}$) and may be disregarded in our kinematic
analysis at the stage of our studies in question.

As can be seen from Fig.~3, the magnitude equation for the proper
motions in right ascension in the magnitude range $13^m-15^m.5$
lies within the range 1 to 2 mas yr$^{-1}$. In this magnitude
range, the variation in differences can be represented by a linear
law with a coefficient of $-0.60\pm0.05$~mas yr$^{-1}$ mag$^{-1}$.

The presence of systematic errors in the coordinates of stars in
the catalogs of early epochs can be one of the main reasons for
the existence of a magnitude equation for faint stars. Studies by
Khrutskaya and Khovritchev (2003, 2004) showed that the YS4.0
catalog (Yellow Sky plates), which was used as the first epoch to
derive the UCAC2 proper motions, could be the the source of the
magnitude equation. As was noted in the description to the UCAC2
catalog, the Yellow Sky plates were remeasured, but the results
have not yet been published.

Platais et al. (1998) showed that there is also a significant
magnitude equation with a coefficient of $\sim1$~mas yr$^{-1}$
mag$^{-1}$ in the published proper motions of NPM stars (blue
plates) $\mu_\alpha\cos\delta$. In contrast to our case, the
magnitude dependence of the Hipparcos$-$NPM differences is
positive in sign.

 \subsection{\it Solution with the Elimination of the Magnitude Equation}

We again solved Eqs. (1) and (2) for UCAC2 magnitudes fainter than
$13^m.0$. The proper motions of stars $\mu_\alpha\cos\delta$ with
declinations $\delta>-5^\circ$ were corrected for the magnitude
equation with a coefficient of $-0.60\pm0.05$ mas yr$^{-1}$
mag$^{-1}$. The magnitude equation for $16^m.5$ stars was assumed
to be zero. Since there is a mixed effect of the first epochs of
the various sources used to derive the proper motions in the
magnitude range $11.5-13^m$, we do not consider these stars.

The derived kinematic parameters are indicated in Fig.~2 by the
solid line and open circles. As can be seen from the figure,
including the magnitude equation affected the three rotation
tensor components. Comparison of Figs.~1 and 2 leads us to
conclude that, first, the magnitude dependence of
 $M_{\scriptscriptstyle21}^{\scriptscriptstyle-}$ (the rotation
around the $z$ axis) for UCAC2 stars corrected for the magnitude
equation is in better agreement with that found from Tycho-2 stars
and, second, the magnitude dependence of
 $M_{\scriptscriptstyle13}^{\scriptscriptstyle-}$ (the rotation
around the $y$ axis) corrected for the magnitude equation is
monotonic, as we expected. We attribute the small jump for stars
fainter than 16m to the fact that a transition to distant stars
occurs. The magnitude dependence of
 $M_{\scriptscriptstyle32}^{\scriptscriptstyle-}$ (the rotation
around the $x$ axis) corrected for the magnitude equation changes
sign, but its absolute value is close to zero. At the same time,
the deformation in the $yz$ plane undergoes noticeable variations
with magnitude with an amplitude of $\approx0.6$~mas yr$^{-1}$;
for the faintest stars,
 $M_{\scriptscriptstyle23}^{\scriptscriptstyle+}\approx0.3\pm0.05$~mas yr$^{-1}$.

Eliminating the magnitude equation for $13^m$ stars changes
  $M_{\scriptscriptstyle21}^{\scriptscriptstyle-}$ by  25$\%$ and
  $M_{\scriptscriptstyle13}^{\scriptscriptstyle-}$ by 100$\%$.

On the whole, we may conclude that the proper motions of UCAC2
stars corrected for the magnitude equation in the magnitude range
$13-15^m$ became systematically closer to the Tycho-2 system.

{
\begin{table*}[t]                                                %% Таблица 1
\caption[]{\small\baselineskip=1.0ex\protect
 Local rotation tensor components $M_{13}^{-}$ and $M_{32}^{-}$
 calculated using distant Tycho-2 (solutions 1--3) and UCAC2 (solutions 4--8)
 stars based on the simultaneous solution of Eqs. (1) and (2).

}
\begin{center}
\begin{tabular}{|c|c|c|c|c|c|c|c|c|c|c|c}\hline
 %&&&&&&&&\\
No. &$N_\star$&$d_W$, pc&$B-V$&${\overline V}$&$J-K_s$
    &${\overline m}$&$M_{13}^{-}$, mas yr$^{-1}$ & $M_{32}^{-}$, mas yr$^{-1}$
\\\hline
%%%% Tycho-2
1 &  89089 & $637\pm 55$ & $\leq0.2$ & 10.17 &&& $-0.488\pm0.046$ & $-0.247\pm0.047$ \\
2 & 228230 & $578\pm 58$ & $  1-1.4$ & 10.32 &&& $-0.455\pm0.033$ & $-0.251\pm0.033$ \\
3 & 130938 & $873\pm100$ & $  >1.4$  & 10.40 &&& $-0.371\pm0.035$ & $-0.266\pm0.035$ \\
%%%% UCAC2
4 & 319129 & $641\pm 60$ &&& $\geq0.5$ & 16.41 & $-0.597\pm0.032$ & $-0.137\pm0.031$ \\
5 & 269675 & $643\pm 63$ &&& $\geq0.5$ & 16.55 & $-0.486\pm0.033$ & $-0.177\pm0.033$ \\
6 & 192852 & $747\pm 76$ &&& $   <0.5$ & 16.41 & $-0.487\pm0.030$ & $-0.014\pm0.030$ \\
7 & 156702 & $774\pm 87$ &&& $   <0.5$ & 16.55 & $-0.486\pm0.032$ & $-0.058\pm0.033$ \\
8 & 211086 & $975\pm127$ &&& $\geq0.8$ & 10.31 & $-0.493\pm0.045$ & $-0.018\pm0.043$ \\
  \hline  %% среднее из 8
Mean &&&&&&& $-0.485\pm0.023$ & $-0.140\pm0.036$ \\
  \hline
%%%% Tycho-2=== Шарлье:
1 &        & $486\pm116$ & $\leq0.2$ & 10.17 &&& $-0.25\pm0.17$ & $-0.27\pm0.17$ \\
2 &        & $470\pm 90$ & $  1-1.4$ & 10.32 &&& $-0.50\pm0.11$ & $-0.17\pm0.11$ \\
3 &        & $749\pm196$ & $  >1.4$  & 10.40 &&& $-0.38\pm0.13$ & $-0.28\pm0.13$ \\
%%%% UCAC2
4 &        & $604\pm126$ &&& $\geq0.5$ & 16.41 & $-0.45\pm0.10$ & $-0.35\pm0.11$ \\
5 &        & $651\pm138$ &&& $\geq0.5$ & 16.55 & $-0.42\pm0.10$ & $-0.39\pm0.10$ \\
6 &        & $744\pm177$ &&& $   <0.5$ & 16.41 & $-0.35\pm0.11$ & $-0.13\pm0.11$ \\
7 &        & $754\pm177$ &&& $   <0.5$ & 16.55 & $-0.37\pm0.10$ & $-0.18\pm0.11$ \\
8 &        & $967\pm292$ &&& $\geq0.8$ & 10.31 & $-0.32\pm0.13$ & $-0.22\pm0.14$ \\
  \hline
Mean &&&&&&& $-0.39\pm0.02$ & $-0.25\pm0.04$ \\
 \hline
\end{tabular}
\end{center}
 \small\baselineskip=1.0ex\protect
 Note: No. is the solution (or sample), $N_\star$ is the number of stars,
 solutions 1--8 in the upper and lower parts of the table were
 obtained, respectively, from individual stars and from the same
 data with a division into Charlier fields.
\end{table*}
}

{
\begin{table*}[t]                                                %% Таблица 2
\caption[]{\small\baselineskip=1.0ex\protect
 Cross-correlation coefficients derived when simultaneously
 solving the system of conditional equations (1) and (2) using individual
 stars (above the diagonal) and by the method with Charlier fields (below the diagonal)
 }
\begin{center}
\begin{tabular}{|c|c|c|c|c|c|c|c|c|c|c|c|c|}\hline
 Parameter   & & 1   & 2   & 3   & 4   & 5   & 6   & 7   & 8    & 9   & 10   & 11   \\\hline
      $X_\odot$& 1& 1.0 & 0.01&-0.01& 0.01& 0.0 &-0.05& 0.02& 0.15 &-0.04&-0.01& 0.0  \\
      $Y_\odot$& 2& 0.0 & 1.0 &-0.01& 0.15& 0.05& 0.0 & 0.20& 0.03 &-0.01&-0.04& 0.0  \\
      $Z_\odot$& 3& 0.0 & 0.0 & 1.0 &-0.01& 0.0 &-0.10& 0.0 & 0.0  & 0.08&-0.02&-0.05 \\
   $M_{21}^{+}$& 4& 0.0 & 0.0 & 0.0 & 1.0 &-0.02& 0.01& 0.03& 0.0  & 0.0 & 0.01& 0.0  \\
   $M_{32}^{-}$& 5& 0.0 & 0.0 & 0.0 & 0.0 & 1.0 & 0.0 &-0.02& 0.02 & 0.0 &-0.62& 0.03 \\
   $M_{13}^{-}$& 6& 0.0 & 0.0 & 0.0 & 0.0 & 0.0 & 1.0 &-0.02& 0.01 & 0.61& 0.0 & 0.0  \\
   $M_{21}^{-}$& 7& 0.0 & 0.0 & 0.0 & 0.0 & 0.0 & 0.0 & 1.0 & 0.01 &-0.02& 0.02& 0.0  \\
$M_{11-22}^{+}$& 8& 0.0 & 0.0 & 0.0 & 0.0 & 0.0 & 0.0 & 0.0 & 1.0  & 0.0 &-0.02& 0.39 \\
   $M_{13}^{+}$& 9& 0.0 & 0.0 & 0.0 & 0.0 & 0.0 & 0.0 & 0.0 & 0.0  & 1.0 & 0.0 & 0.03 \\
   $M_{23}^{+}$&10& 0.0 & 0.0 & 0.0 & 0.0 & 0.0 & 0.0 & 0.0 & 0.0  & 0.0 & 1.0 &-0.01 \\
$M_{33-22}^{+}$&11& 0.0 & 0.0 & 0.0 & 0.0 & 0.0 & 0.0 & 0.0 & 0.50 & 0.0 & 0.0 & 1.0 \\
 \hline
\end{tabular}
\end{center}
 \small\baselineskip=1.0ex\protect
 Note: The coefficients are given for sample No. 3 from Table 1.
\end{table*}
}

\subsection{\it Estimating the Group Distances}

To estimate the distances to stars, we use a statistical method
(Olling and Dehnen 2003). As the peculiar solar velocity relative
to the local standard of rest, we take the values from Dehnen and
Binney (1998):
 $(U_\odot,V_\odot,W_\odot)=(10.00,5.25,7.17)\pm(0.36,0.62,0.38)$~km s$^{-1}$.
We calculate the parallaxes using the formulas
 $$
 \pi_U={{4.74\cdot X_\odot}\over U_\odot},  ~~~
 \pi_W={{4.74\cdot Z_\odot}\over W_\odot},
 $$
where $X_\odot$ and $Z_\odot$ are the stellar group velocity
components that we found by solving Eqs. (1) and (2), in mas
yr$^{-1}$. Since the $Y_\odot$ component is distorted appreciably
by the asymmetric drift (Dehnen and Binney 1998), this component
is not used to determine the group parallaxes. We find the
distance $d$ from the relation $d = 1/\pi$. In this case, the
distance error can be estimated from the relation
 $$
 e_d=\biggl({e_\pi\over \pi}\biggr)\cdot d,
 $$
we estimate $e_\pi$ for the motion along the $z$ axis as
 $$
 e_\pi=4.74\cdot\sqrt
 {
   \biggl( {{e_{W_\odot} \cdot Z_\odot} \over {W_\odot}^2} \biggr)^2
 + \biggl( {                e_{Z_\odot} \over {W_\odot}} \biggr)^2
 }.
 $$
A similar relation can be derived for the motion along the $x$
axis. Practice shows (Dehnen and Binney 1998; Olling and Dehnen
2003) that the $z$ components of the solar velocity are more
stable than the $x$ velocity components for stars of various
Galactic subsystems; therefore, the parallaxes $\pi_W$ are
preferred to $\pi_U$.

First, we apply the method to the Hipparcos catalog. The division
into groups is performed using trigonometric parallaxes of the
catalog with $e_\pi/\pi<1$. In Fig.~4, our statistical distances
to stars are plotted against trigonometric distances. We see from
the figure that there are no significant systematic distortions in
the statistical distances and that the errors in the statistical
distances at distances of more than 200~pc are smaller than those
in the trigonometric distances. In particular, we estimated the
mean distance for 4216 Hipparcos stars with negative trigonometric
parallaxes and obtained the following values:
 $d_U= 838\pm244$~pc and
 $d_W= 716\pm211$~pc.
This confirms our assumption that such stars are, on average, far
from the Sun (Bobylev 2004a).

Next, we use the method to analyze UCAC2 stars. The proper motions
of northern-sky stars fainter than $13^m.0$ were corrected for the
magnitude equation found. In Fig.~5, the statistical distances to
stars calculated using the parallaxes $\pi_W$ are plotted against
the distances found using $\pi_U$. As can be seen from the figure,
there is a difference in the distances $1/\pi_W$ and $1/\pi_U$ at
$\sim$600~pc that gradually disappears near 700~pc. This
difference corresponds to the stars in the magnitude range
$12-13^m$ and stems from the fact that the magnitude equation
(which was not compensated for in this range) affects the
determination of the $X_\odot$ velocity component. We see that
stars fainter than $16^m.0$ are, on average, farther than
$650-700$~pc, with the error in the group distances being no
larger than $14\%$. The distance found shows that these stars are
located, on average, outside the local (Orion) spiral arm or the
Local system of stars. This gives reason to consider the rotation
around the Galactic $y$ axis found using the faintest stars
(Fig.~2) as a residual rotation of the ICRS/Tycho-2 system
relative to the extragalactic coordinate system.

 \subsection{\it Estimating the Residual Rotation of the ICRS/Tycho-2 System}

{\bf Results of the simultaneous solution of Eqs.(1) and (2).}
 Let us estimate the limit to which
 $M_{\scriptscriptstyle13}^{\scriptscriptstyle-}$ tends as the distance to stars
increases. We proceed from the fact that, on the whole, distant
stars of various spectral types belonging to various Galactic
components may not have any rotation around the $y$ axis. At the
same time, we consider the presence of a significant rotation as a
residual rotation of the ICRS/Tycho-2 system relative to the
extragalactic coordinate system.

To solve the formulated problem, we use almost all of the Tycho-2
stars fainter than $8^m.5$ and all of the UCAC2 stars fainter than
$16^m.34$ (about one million stars). We gave preference to the
Tycho-2 catalog in the range of bright stars  because of its full
sky coverage. In the UCAC2 catalog, we selected the range with a
small magnitude equation that we disregard.

To extend the distance scale where possible, we rely on a
well-known effect related to the fact that stars of different
brightnesses and colors are, on average, located at different
distances. As a result, we divided the Tycho-2 stars into two
parts in magnitude with the boundary $V=11^m.25$, into bright and
faint ones, and then divided each part into seven parts in $B-V$
color in the following ranges: $<$0.2, 0.2--0.4, 0.4--0.6,
0.6--0.8, 0.8--1, 1--1.4, $>$1.4.

We use only those stars from the UCAC2 catalog for which the
photometric magnitudes from the 2MASS survey are given, $J,H,K_s$.
In this case, it is important to ensure that the stars cover the
entire sky. As a result, we divided the stars into four parts in
both magnitude and $J-K_s$ color (with the $0^m.5$ boundary).

The results obtained are presented in Fig.~6 and Table~1. As we
see from Fig.~6, the rotation,
 $M_{\scriptscriptstyle13}^{\scriptscriptstyle-}$, and deformation,
 $M_{\scriptscriptstyle13}^{\scriptscriptstyle+}$, tensor
components undergo noticeable and coherent variations with
increasing distance. Analysis of the plots leads us to conclude
that there is an actual rotation of the stars confirmed by
deformations up to a distance of $\sim$600~pc. This effect stems
from the fact that the stars belong to the Local system. This
boundary is most distinct in Fig.~6b, where
$M_{\scriptscriptstyle13}^{\scriptscriptstyle+}$ decreases sharply
to zero at distances larger than 600~pc.

Table~1 lists the following parameters: the solution (or sample)
number, the number of stars, the distance $d_W$, the $B-V$ and $V$
ranges for Tycho-2 stars, $J-K_s$ and the mean magnitude (U2R) for
UCAC2 stars; the last two columns give the derived components of
the vector of rotation around the y and $x$ axes. Solution
No.~$1-3$ pertain to bright ($8^m.5-11^m.25$) Tycho-2 stars;
solution No.~$4-8$ were obtained from UCAC2 stars. The solutions
obtained using individual stars and the same data, but with a
division into Charlier fields, are given in the upper and lower
parts of the table, respectively. The maximum number of fields is
432 for Tycho-2 and 367 for UCAC2. For each method, the table
gives the corresponding weighted mean parameters
$M_{\scriptscriptstyle13}^{\scriptscriptstyle-}$ and
$M_{\scriptscriptstyle13}^{\scriptscriptstyle+}$.

Table~2 gives the correlation coefficients obtained when
simultaneously solving the system of conditional equations (1) and
(2) for sample No.~3. As we see, applying the method with fields
reduces significantly the correlation between the unknowns
($M_{\scriptscriptstyle13}^{\scriptscriptstyle-}$)--
($M_{\scriptscriptstyle13}^{\scriptscriptstyle+}$)
 and
($M_{\scriptscriptstyle32}^{\scriptscriptstyle-}$)--
($M_{\scriptscriptstyle23}^{\scriptscriptstyle+}$) being
determined.

As can be seen from the upper part of Table 1,
$M_{\scriptscriptstyle13}^{\scriptscriptstyle-}$ obtained from
bright Tycho-2 stars with $B-V>1.4$ (solution No.~3) is the
minimum one (in absolute value). Together with distant M-type
giants, sample No.~3 also includes nearby main-sequence dwarfs.
Giants can be separated more reliably from dwarfs using 2MASS
photometry (see, e.g., Majewski et al. 2003; Babusiaux and Gilmore
2005). We decided to check this result using bright (U2Rmag:
$8^m.0-11^m.17$) stars from the UCAC2 catalog, where the magnitude
equation is negligible (solution No.~8). For this purpose, we
selected 211 086 stars according to the criterion $J-K_s>0.8$; the
stars lie in the $K_s$ magnitude range $3-7^m$ (the mean is
$6^m.75$).We chose the boundary $J-K_s>0.8$ empirically so as to
have a representative sample. Most importantly, $\approx50\%$ of
the stars lie outside the zone of the Galactic plane, i.e., have
$|b|>10^\circ$. We have the opposite picture for other distance
indicators, distant O and B stars (solution No.~1), since almost
all of them are distributed in the Galactic equatorial plane,
which, in particular, reduces the reliability of the
$M_{\scriptscriptstyle13}^{\scriptscriptstyle-}$ estimate. For
solution No.~1, the correlation coefficients between the unknowns
($M_{\scriptscriptstyle13}^{\scriptscriptstyle-}$)-($M_{\scriptscriptstyle13}^{\scriptscriptstyle+}$)
and
($M_{\scriptscriptstyle32}^{\scriptscriptstyle-}$)-($M_{\scriptscriptstyle23}^{\scriptscriptstyle+}$)
being determined are $\approx0.85$. Comparison of our estimates of
the distances $d_U$ and $d_W$ for bright UCAC2 stars with the
photometric distances from Babusiaux and Gilmore (2005) for the
$(K_s)-(J-K_s)$ plane leads us to conclude that the estimate of
$d\approx1$~kpc for our sample No.~8 of stars is quite realistic.

Applying the method with fields showed that solution No.~1 is
least reliable. This is because the number of stars outside the
zone of the Galactic plane ($b > 10^\circ$) is $8-10$; in all the
remaining solutions, the number of stars in such fields is an
order of magnitude larger. Solutions No.~3 and No.~8 are most
reliable. In the strict sense, they are not independent, since the
Hipparcos and Tycho-2 catalogs were used (Zacharias et al. 2004)
as the first epoch to determine the proper motions of bright (to
$U2R\approx12^m.5$) UCAC2 stars.

The mean Oort constants were found from all of the UCAC2 stars
fainter than $16^m.34$ using the method of fields to be
 $A = 12.10\pm0.58$ km s$^{-1}$ kpc$^{-1}$ and
 $B =-10.63\pm0.46$ km s$^{-1}$ kpc$^{-1}$.

The estimate of
 $M_{\scriptscriptstyle13}^{\scriptscriptstyle-}=-0.37\pm0.04$ mas yr$^{-1}$
obtained from solution No.~3 in the upper part of Table 1 is in
good agreement with
 $M_{\scriptscriptstyle13}^{\scriptscriptstyle-}=-0.38\pm0.13$ mas yr$^{-1}$
calculated by the method of fields (the lower part of Table 1) and
with the mean
 $M_{\scriptscriptstyle13}^{\scriptscriptstyle-}=-0.39\pm0.02$ mas yr$^{-1}$
calculated using all of the data in the lower part of Table 1.

The mean
 $M_{\scriptscriptstyle13}^{\scriptscriptstyle-}$
calculated using the data in the lower part of Table 1 is fairly
stable, depending on the constraint imposed on $|\mu_t|$. Above,
we used the criterion $|\mu_t| < 300$ mas yr$^{-1}$ (in two
iterations, with analysis of the residuals using the $3\sigma$
criterion), which allows nearby stars with high velocities to be
rejected. We calculated the lower part of Table 1 for various
constraints on $|\mu_t|$, from 300 to 25 mas yr$^{-1}$. The
constraint $|\mu_t|<25$ mas yr$^{-1}$ implies that at a mean
stellar distance of 700 pc, the maximum stellar space velocity is
110 km s$^{-1}$, i.e., only disk stars fall into the sample. The
means (sample No. 1 was not considered because of the small number
of stars at high latitudes remaining under these constraints) are
$$
\displaylines{\hfill
 M_{\scriptscriptstyle13}^{\scriptscriptstyle-}=-0.33\pm0.02~\hbox {mas yr$^{-1}$},\hfill\llap(4)\cr\hfill
 M_{\scriptscriptstyle13}^{\scriptscriptstyle+}=+0.05\pm0.06~\hbox {mas yr$^{-1}$},\hfill\cr\hfill
 M_{\scriptscriptstyle32}^{\scriptscriptstyle-}=-0.26\pm0.04~\hbox {mas yr$^{-1}$},\hfill\cr\hfill
 M_{\scriptscriptstyle23}^{\scriptscriptstyle+}=+0.03\pm0.04~\hbox {mas yr$^{-1}$},\hfill\cr\hfill
 |\mu_t|<25~\hbox {mas yr$^{-1}$}.\hfill\cr
 }
$$
%%
%%
%%
%%%%%%%%%%%%%%%%%%%%%%%%%%%%%%%%%%%%%%%%%%%%%%
{
\begin{table*}[t]                                                %% Таблица 3
\caption[]{\small\baselineskip=1.0ex\protect
 Kinematic parameters calculated from distant Tycho-2
 stars using only Eq.~(2).
 }
\begin{center}
\begin{tabular}{|c|c|c|c|c|c|c|c|c|c}\hline
 %&&&&&&\\
${\overline V}$&$d_W$, pc&$A$, km s$^{-1}$ kpc$^{-1}$&$M_{13}^{-}$, mas yr$^{-1}$&$M_{13}^{+}$, mas yr$^{-1}$&$M_{32}^{-}$, mas yr$^{-1}$&$M_{23}^{+}$, mas yr$^{-1}$\\
  \hline
10.32&$572\pm 55$&$13.8\pm0.4$&$-0.30\pm0.05$ & $-0.09\pm0.06$ & $-0.20\pm0.05$ & $-0.05\pm0.06$\\
10.40&$856\pm 94$&$13.9\pm0.5$&$-0.27\pm0.06$ & $-0.15\pm0.07$ & $-0.20\pm0.06$ & $-0.07\pm0.07$\\
  \hline
10.32&$484\pm 82$&$12.5\pm1.0$&$-0.33\pm0.09$ & $-0.36\pm0.17$ & $-0.19\pm0.09$ & $+0.16\pm0.17$\\
10.40&$714\pm165$&$11.1\pm1.2$&$-0.29\pm0.11$ & $-0.37\pm0.21$ & $-0.22\pm0.11$ & $+0.04\pm0.21$\\
 \hline
\end{tabular}
\end{center}
 \small\baselineskip=1.0ex\protect
 Note:
 The solutions obtained from individual stars and from the same data
 with a division into Charlier fields are given in the upper and
 lower parts of the table, respectively.
 \end{table*}
}
%%%%%%%%%%%%%%%%%%%%%%%%%%%%%%%%%%%%%%%%%%%%%%
%%%%%%%%%%%%%%%%%%%%%%%%%%%%%%%%%%%%%%%%%%%%%%
{
\begin{table*}[t]                                                %% Таблица 4
\caption[]{\small\baselineskip=1.0ex\protect
 Kinematic parameters calculated from distant UCAC2
 stars using only Eq.~(2).
 }
\begin{center}
\begin{tabular}{|c|c|c|c|c|c|c|c|c|c}\hline
 %&&&&&&\\
${\overline m}$&$d_W$, pc&$A$,km s$^{-1}$ kpc$^{-1}$&$M_{13}^{-}$, mas yr$^{-1}$&$M_{13}^{+}$, mas yr$^{-1}$&$M_{32}^{-}$, mas yr$^{-1}$&$M_{23}^{+}$, mas yr$^{-1}$\\
  \hline
16.41&$ 731\pm 65$& $12.3\pm0.4$ &$-0.41\pm0.04$ & $0.05\pm0.04$ & $-0.14\pm0.04$ & $-0.28\pm0.11$\\
16.55&$ 727\pm 68$& $13.4\pm0.4$ &$-0.42\pm0.04$ & $0.06\pm0.05$ & $-0.20\pm0.04$ & $-0.31\pm0.06$\\
10.31&$1122\pm128$& $13.9\pm0.4$ &$-0.17\pm0.05$ & $0.26\pm0.06$ & $-0.17\pm0.06$ & $+0.19\pm0.06$\\
  \hline
16.41&$ 701\pm142$& $10.9\pm1.4$ &$-0.32\pm0.10$ & $0.26\pm0.15$ & $-0.16\pm0.11$ & $-0.12\pm0.15$\\
16.55&$ 719\pm149$& $11.2\pm1.4$ &$-0.29\pm0.11$ & $0.35\pm0.15$ & $-0.26\pm0.11$ & $-0.15\pm0.15$\\
10.31&$1184\pm274$& $12.2\pm1.0$ &$-0.27\pm0.08$ & $0.05\pm0.11$ & $-0.12\pm0.09$ & $+0.18\pm0.11$\\
 \hline
\end{tabular}
\end{center}
 \small\baselineskip=1.0ex\protect
 Note:
 The solutions obtained from individual stars and from the same data
 with a division into Charlier fields are given in the upper and
 lower parts of the table, respectively.
\end{table*}
}
%%%%%%%%%%%%%%%%%%%%%%%%%%%%%%%%%%%%%%%%%%%%%%
%%
%%
%%
%%
{\bf Results of the solution of only Eq. (2).} Tables 3 and 4 give
the kinematic parameters that were determined by solving only Eq.
(2). We used two approaches: the solution based on individual
stars (the upper parts of Tables 3 and 4). For the Tycho-2 catalog
(Table 3), we used the samples of stars corresponding to solutions
No.~2 and 3, whose parameters are listed in Table. 1. We imposed
the constraint
 $|\mu_t| < 100$ mas yr$^{-1}$
on the absolute value of the tangential stellar velocity. Using
this constraint reduced considerably the random errors of the
unknowns in the method with fields and affected noticeably sample
no.~8: the mean distance to the stars increased to
$d_W\approx1100$~pc.

For the UCAC2 catalog (Table 4), we used the faintest stars (the
stars whose parameters are listed in Table 1 for solutions nos. 4
and 6 as well as for solutions nos. 5 and 7 were combined into two
groups); about 450 000 stars were used in each magnitude range.

The results of solving Eq. (2) obtained by the two methods show
that the Oort constant $A$ and the local rotation tensor
components are determined reliably and confirm the results that we
obtained by simultaneously solving the system of conditional
equations (1) and (2).

Table 5 gives the correlation coefficients derived when solving
Eq. (2) for sample no. 3. As we see from Table 5, applying the
method with fields reduces the correlation between the unknowns
($M_{\scriptscriptstyle13}^{\scriptscriptstyle-}$)-($M_{\scriptscriptstyle13}^{\scriptscriptstyle+}$)
and
($M_{\scriptscriptstyle32}^{\scriptscriptstyle-}$)-($M_{\scriptscriptstyle23}^{\scriptscriptstyle+}$)
being determined by an order of magnitude. Comparison of Tables 5
and 2 leads us to conclude that the correlation between the above
unknowns is lower when Eqs. (1) and (2) are solved simultaneously;
therefore, the simultaneous solution is preferred.

In our model, we did not abandon the unknowns
$(M_{\scriptscriptstyle11}^{\scriptscriptstyle+}-
M_{\scriptscriptstyle22}^{\scriptscriptstyle+})$ and
$(M_{\scriptscriptstyle33}^{\scriptscriptstyle+}-
M_{\scriptscriptstyle22}^{\scriptscriptstyle+})$. The kinematic
meaning of the first difference is that
$0.5(M_{\scriptscriptstyle11}^{\scriptscriptstyle+}-
M_{\scriptscriptstyle22}^{\scriptscriptstyle+})=C$, where $C$ is
the Oort constant. We can determine the vertex deviation $l_{xy}$
in the $xy$ plane from the relation $\tan 2l_{xy}=-C/A$ by
analyzing the Oort constants $A$ and $C$. Our results show that
the vertex deviation for the most distant stars under
consideration (sample no. 8) differs significantly from zero; it
is
 $l_{xy}=7\pm1^\circ$ and, for sample no. 3, $l_{xy}=6\pm2^\circ$.

%%%%%%%%%%%%%%%%%%%%%%%%%%%%%%%%%%%%%%%%%%%%%%
{
\begin{table*}[t]                                                %% Таблица 5
\caption[]{\small\baselineskip=1.0ex\protect
 Cross-correlation coefficients obtained when solving Eq.(2) using individual stars
 (above the diagonal) and by method with Charlier fields (below the diagonal)
 }
\begin{center}
\begin{tabular}{|c|c|c|c|c|c|c|c|c|c|c|c|}\hline
   Parameter   && 1   & 2   & 3   & 4   & 5   & 6   & 7   & 8    & 9   & 10    \\\hline
      $X_\odot$& 1& 1.0 & 0.0 &-0.02& 0.01&-0.01&-0.06& 0.12&-0.04& 0.0 &-0.01 \\
      $Y_\odot$& 2& 0.0 & 1.0 &-0.02& 0.14& 0.05& 0.01&-0.02& 0.0 &-0.03&-0.03 \\
      $Z_\odot$& 3& 0.0 & 0.0 & 1.0 &-0.02& 0.01&-0.06& 0.0 & 0.04&-0.02&-0.03 \\
   $M_{21}^{+}$& 4& 0.0 & 0.0 & 0.0 & 1.0 & 0.0 & 0.03& 0.01& 0.02&-0.01& 0.0  \\
   $M_{32}^{-}$& 5& 0.0 & 0.0 & 0.0 & 0.0 & 1.0 & 0.0 & 0.01& 0.0 &-0.89& 0.03 \\
   $M_{13}^{-}$& 6& 0.0 & 0.0 & 0.0 & 0.0 & 0.0 & 1.0 &-0.02& 0.88& 0.0 &-0.01 \\
$M_{11-22}^{+}$& 7& 0.0 & 0.0 & 0.0 & 0.0 & 0.0 & 0.0 & 1.0 &-0.04&-0.01& 0.81 \\
   $M_{13}^{+}$& 8& 0.0 & 0.0 & 0.0 & 0.0 & 0.0 & 0.02& 0.0 & 1.0 & 0.0 &-0.01 \\
   $M_{23}^{+}$& 9& 0.0 & 0.0 & 0.0 & 0.0 &-0.02& 0.0 & 0.0 & 0.0 & 1.0 &-0.02 \\
$M_{33-22}^{+}$&10& 0.0 & 0.0 & 0.0 & 0.0 & 0.0 & 0.0 & 0.82& 0.0 & 0.0 & 1.0  \\
 \hline
\end{tabular}
\end{center}
 \small\baselineskip=1.0ex\protect
 Note: The coefficients are given for sample No. 3 from Table 1.
\end{table*}
}
%%%%%%%%%%%%%%%%%%%%%%%%%%%%%%%%%%%%%%%%%%%%%%

\section{DISCUSSION}

On the whole, we may conclude that the estimate
 $M_{\scriptscriptstyle13}^{\scriptscriptstyle-}=-0.37\pm0.04$ mas yr$^{-1}$
can be interpreted as a residual rotation of the ICRS/Tycho-2
system relative to the inertial reference frame. This value is in
good agreement with
 $M_{\scriptscriptstyle13}^{\scriptscriptstyle-}=-0.36\pm0.09$ mas yr$^{-1}$
that we obtained by analyzing distant Hipparcos stars (Bobylev
2004a) using trigonometric parallaxes to analyze the residual
rotation of the ICRS/HIPPARCOS system as a function of distance.
Previously (Bobylev 2004b), based on an independent astrometric
method, we found the following equatorial components of the vector
of residual rotation of the ICRS/HIPPARCOS system relative to the
inertial reference frame:
 $\omega_x= +0.04\pm0.15$ mas yr$^{-1}$,
 $\omega_y= +0.18\pm0.12$ mas yr$^{-1}$, and
 $\omega_z= -0.35\pm0.09$ mas yr$^{-1}$.
Transforming the $\omega_z$ component to the Galactic coordinate
system yields
 $M_y=M_{\scriptscriptstyle13}^{\scriptscriptstyle-}=-0.26\pm0.07$ mas yr$^{-1}$,
which is also in agreement with the above estimate.

Inaccurate realization of the ICRS/Hipparcos system, in other
words, a residual rotation of the ICRS/Hipparcos system relative
to ICRF, can be one of the main causes of the rotation found (Ma
et al. 1998). Since a linear model of solid-body rotation
(Kovalevsky et al. 1997) was used to link the ICRS/Hippacros
system to ICRF, we believe it appropriate to use the linear
Ogorodnikov-Milne model to solve the formulated problem.

The best solution of the problem of controlling the inertiality of
the ICRS/Tycho-2 system may be associated with the use of images
for extragalactic objects in the UCAC2 catalog. As follows from
the description to the electronic version of UCAC2, the measured
images of star like extragalactic objects are present. In the
catalog, they are not marked in any way and are not identified;
their total number is unknown. As follows from our work, for the
random errors in the sought-for parameters to be
 $\sim$0.05 mas yr$^{-1}$,
we must have $\sim$100000 objects.

On the other hand, we see from Fig. 6 that the variations in the
rotation and deformation parameters with distance are nonlinear in
pattern. This is attributable to peculiarities of the structure of
the Local system of stars and larger-scale Galactic structures
(the spiral arms and the disk warp). The possibilities for a more
detailed study of these peculiarities will emerge soon, when the
next version of the UCAC2 catalog, providing full coverage of the
celestial sphere, will be published. In this case, the magnitude
equation should be completely eliminated from the proper motions
of its stars.

Our results have a bearing on the problem of studying the Galactic
disk warp (Miyamoto and S\^oma 1993; Miyamoto et al. 1993b;
Miyamoto and Zhu 1998). Since O--B5 stars concentrated near the
Galactic plane are mainly analyzed in these papers, the
correlation coefficients between the parameters with subscript 3
are close to 0.95. To reduce the  correlations, the
Ogorodnikov-Milne model is simplified by making the assumptions
(in our notation)
 $M_{\scriptscriptstyle13}^{\scriptscriptstyle+}=
 -M_{\scriptscriptstyle13}^{\scriptscriptstyle-}$ and
 $M_{\scriptscriptstyle32}^{\scriptscriptstyle-}=
  M_{\scriptscriptstyle23}^{\scriptscriptstyle+}$,
suggesting that
 ${\strut\displaystyle \partial\displaystyle  V_R     /\displaystyle\partial\displaystyle z}=
  {\strut\displaystyle \partial\displaystyle  V_\theta/\displaystyle\partial\displaystyle
  z}=0$.
Under these assumptions, the above terms disappear from Eq. (1).
In this case, the key equation is Eq. (2), which allows us to
determine the quantities
 $ 2M_{\scriptscriptstyle13}^{\scriptscriptstyle+}=
  -2M_{\scriptscriptstyle13}^{\scriptscriptstyle-}=
 {\strut\displaystyle \partial\displaystyle  V_z/\displaystyle\partial\displaystyle R}$
and
 $ 2M_{\scriptscriptstyle32}^{\scriptscriptstyle-}=
   2M_{\scriptscriptstyle23}^{\scriptscriptstyle+}=
 ({\strut\displaystyle1/\displaystyle R})
 ({\strut \displaystyle\partial\displaystyle V_z/\displaystyle\partial\displaystyle\theta})$.
The other model of these authors is constructed by assuming that
 $M_{\scriptscriptstyle13}^{\scriptscriptstyle-}=
  M_{\scriptscriptstyle13}^{\scriptscriptstyle+}=0$,
 $M_{\scriptscriptstyle32}^{\scriptscriptstyle-}=
  M_{\scriptscriptstyle23}^{\scriptscriptstyle+}$.
In this case, only one quantity,
$2M_{\scriptscriptstyle32}^{\scriptscriptstyle-}$, is determined
from the solution of the system of conditional equations (1) and
(2). As we showed, using the method with fields, which is almost
free from any correlations between the unknowns
($M_{\scriptscriptstyle13}^{\scriptscriptstyle-}$)--($M_{\scriptscriptstyle13}^{\scriptscriptstyle+}$)
 and
($M_{\scriptscriptstyle32}^{\scriptscriptstyle-}$)--($M_{\scriptscriptstyle23}^{\scriptscriptstyle+}$),
allows the problem to be considered without any preliminary
assumptions. The parameters in Tables 3 and 4 show that
$M_{\scriptscriptstyle13}^{\scriptscriptstyle+}$,
$M_{\scriptscriptstyle32}^{\scriptscriptstyle-}$ and
$M_{\scriptscriptstyle23}^{\scriptscriptstyle+}$ have
statistically significant differences, depending on the method of
analysis and the sample.

Using Galactic stars to solve our problem has a number of
peculiarities related to the structure of the Galaxy. Let us
consider some of them.

(1) The $yz$ plane, rotation around the x axis. Chiba and Beers
(2000) showed that old (metal-poor) thick-disk stars lag well
behind the circular Galactic rotation velocity,
 $dV_\theta/d|z|=-30\pm3$ km s$^{-1}$ kpc$^{-1}$;
for halo stars, this gradient is almost twice as large. This
quantity expressed in mas yr$^{-1}$ is $30/4.74=6.3$~mas
yr$^{-1}$; it can have an effect if the distribution of sample
stars is asymmetric in $z$. As can be seen from Eqs. (3), the
gradient
 ${\strut\displaystyle\partial\displaystyle V_\theta/\displaystyle\partial\displaystyle z}$
enters into the parameters
 $M_{\scriptscriptstyle23}^{\scriptscriptstyle+}$ and
$M_{\scriptscriptstyle32}^{\scriptscriptstyle-}$. Applying the
method with fields and using the constraint on $|\mu_t|$ (solution
(4)) allows this effect to be minimized. The presence of an
appreciable component
 $M_{\scriptscriptstyle32}^{\scriptscriptstyle-}$ (solution (4))
should be considered as a manifestation of a particular effect in
the actual kinematics of stars, since we consider
$M_{\scriptscriptstyle13}^{\scriptscriptstyle-}$ as the main
component in the problem of controlling the inertiality of the
ICRS system as a residual rotation (Bobylev 2004b). The
relationship between distant giants and kinematics of the warped
Galactic disk may be such an effect. However, a detailed analysis
of this problem is beyond the scope of this paper.

(2) The $xy$ plane, rotation around the $z$ axis. The linear
Ogorodnikov-Milne model we used allows a parabolic fit to the
Galactic rotation curve ($V_\theta(R)$) near $R_\circ$ to be
obtained. Analysis of the currently available data shows that the
Galactic rotation curve at the solar distance is flat (Popova and
Loktin 2005; Avedisova 2005; Zabolotskikh et al. 2002). Previously
(Bobylev 2004c), we determined the following parameters of the
angular velocity ($\omega(R)$) of Galactic rotation by analyzing
distant ($d=1$~kpc) O and B stars:
  $\omega_\circ  =-28.0\pm0.6 $ km s$^{-1}$ kpc$^{-1}$,
  $\omega'_\circ =+4.17\pm0.14$ km s$^{-1}$ kpc$^{-2}$,
  $\omega''_\circ=-0.81\pm0.12$ km s$^{-1}$ kpc$^{-3}$,
 $R_\circ=7.1$ kpc.
Based on these data and assuming that $R-R_\circ=1$~kpc, we can
see that the effect of the second derivative
 $|0.5(R-R_\circ)^2\omega''_\circ|=0.4$ km s$^{-1}$ kpc$^{-1}$ is not significant, is
comparable to the random errors in the Oort constant $A$ and $B$.
Thus, disregarding the second derivative $\omega''_\circ$ gives a
contribution to the amplitude of the waves that we see in the
upper plots of Figs. 1 and 2 (
$M_{\scriptscriptstyle12}^{\scriptscriptstyle+}$ and
 $M_{\scriptscriptstyle21}^{\scriptscriptstyle-}$) of $\sim$0.1 mas yr$^{-1}$,
which is considerably smaller than the amplitude observed in the
plots.

\section{CONCLUSIONS}

Our analysis showed that the proper motions of UCAC2 stars are
distorted by a complex magnitude equation. Thus, for example, no
noticeable effect of the magnitude equation was found in the
proper motions of stars brighter than $\approx11^m$, but this
effect is significant in the magnitude range $12-15^m$. We showed
that the magnitude equation has the greatest effect in
$\mu_\alpha\cos\delta$ of northern-sky stars. This is because
different sources for the southern and northern skies were used as
the first epoch in deriving the proper motions of UCAC2 stars. We
determined the coefficient of the magnitude equation for northern
sky stars fainter than $13^m$ by comparing the proper motions of
UCAC2 and PUL3SE stars, $-0.60\pm0.05$ mas yr$^{-1}$ mag$^{-1}$.
We associate this magnitude equation with the effect of the NPM
catalog, Yellow Sky plates. The derived magnitude equation mainly
affects the determination of the rotation tensor components. Its
elimination leads to a $25\%$ change in the rotation around the z
axis (the Oort constant B) and to a $100\%$ change in the rotation
around the $y$ axis for $13^m$ stars.

The method of statistical parallaxes was used to estimate the
distances to stars with random errors no larger than $14\%$. The
linear solar velocity relative to the local standard of rest,
which is well determined for the local centroid ($d\approx150$
pc), was used as a reference. This could affect the distance
estimates for distant stars.

The method with fields was shown to reduce the correlations
between the unknowns
 ($M_{\scriptscriptstyle13}^{\scriptscriptstyle-}$)--($M_{\scriptscriptstyle13}^{\scriptscriptstyle+}$)
and
($M_{\scriptscriptstyle32}^{\scriptscriptstyle-}$)--($M_{\scriptscriptstyle23}^{\scriptscriptstyle+}$),
being determined by an order of magnitude. When Eqs. (1) and (2)
are solved simultaneously, the correlations between these unknowns
are lower (than those for the solution of only Eq. (2));
therefore, the simultaneous solution is preferred.

The mean Oort constants estimated from Tycho-2 stars
($8^m.5-11^m.25$) are
 $A = 13.88\pm0.98$ km s$^{-1}$ kpc$^{-1}$ and
 $B =-12.13\pm0.63$ km s$^{-1}$ kpc$^{-1}$.
The analogous means that we estimated from bright ($8^m-11^m.17$)
UCAC2 stars have no statistically significant differences. The
mean Oort constants estimated from UCAC2 stars fainter than
$16^m.34$ are
 $A= 12.10\pm0.58$ km s$^{-1}$ kpc$^{-1}$ and
 $B=-10.63\pm0.46$ km s$^{-1}$ kpc$^{-1}$.

We established that the two components of the Ogorodnikov-Milne
model that describe the rotation around the Galactic y axis and
the deformation in the xz plane depend on heliocentric distance.
For the nearest stars ($d\approx200$~pc), these parameters are
$-1.0\pm0.1$~mas yr$^{-1}$ and decrease, on average, by a factor
of 2 at distances of $\sim$600 пк. We associate this with
kinematic peculiarities of the Local system of stars. For more
distant stars, there are, on average, no statistically significant
deformation in the xz plane.

For distant Tycho-2/UCAC2 stars at heliocentric distances of
$\approx$900~pc, the mean rotation around the Galactic y axis is
$-0.37\pm0.04$~mas yr$^{-1}$, which we interpret as a residual
rotation of the ICRS/Tycho-2 system relative to the inertial
reference frame.

\section{ACKNOWLEDGMENTS}

We wish to thank the referees for helpful remarks that contributed
to an improvement of the paper. This work was supported by the
Russian Foundation for Basic Research (project no. 05--02--17047).

\section{REFERENCES}

{\small\noindent
 1. V.S. Avedisova, Astron. Zh. {\bf 82}, 488 (2005) [Astron.
Rep. {\bf 49}, 435 (2005)].

\noindent 2. C. Babusiaux and G. Gilmore, Mon. Not. R. Astron.
Soc. {\bf 358}, 1309 (2005).

\noindent 3. V.V. Bobylev, Pis'ma Astron. Zh. {\bf 30}, 289
(2004a) [Astron. Lett. {\bf 30}, 251 (2004a)].

\noindent 4. V.V. Bobylev, Pis'ma Astron. Zh. {\bf 30}, 930
(2004b) [Astron. Lett. {\bf 30}, 848 (2004b)].

\noindent 5. V.V. Bobylev, Pis'ma Astron. Zh. {\bf 30}, 185
(2004c) [Astron. Lett. {\bf 30}, 159 (2004c)].

\noindent 6. V.V. Bobylev, N.M. Bronnikova, and N.A. Shakht,
Pis'ma Astron. Zh. {\bf 30}, 519 (2004) [Astron. Lett. {\bf 30},
469 (2004)].

\noindent
7. Carlsberg Meridian Catalogue La Palma 13 (Copenhagen
Univ. Obs., Royal Greenwich Obs., Real Inst. y Obs. de la Armada
en San Fernando, 2003).

\noindent 8. M. Chiba and T.C. Beers, Astron. J. {\bf 119}, 2843
(2000).

\noindent 9. S.V.M. Clube, Mon. Not. R. Astron. Soc. {\bf 159},
289 (1972).

\noindent 10. S.V.M. Clube, Mon. Not. R. Astron. Soc. {\bf 161},
445 (1973).

\noindent 11. W. Dehnen and J.J. Binney, Mon. Not. R. Astron. Soc.
{\bf 298}, 387 (1998).

\noindent 12. E. H{\o}g, A. Kuzmin, U. Bastian, et al., Astron.
Astrophys. {\bf 333}, L65 (1998).

\noindent 13. E. H{\o}g, C. Fabricius, V.V. Makarov, et al.,
Astron. Astrophys. {\bf 355}, L27 (2000).

\noindent 14. E.V. Khrutskaya and M. Yu. Khovritchev, {\it
JOURN\'EES 2003}, Ed. by N. Capitaine (Obs. De Paris, Paris,
2003), p. 79.

\noindent 15. E.V. Khrutskaya, M. Yu. Khovritchev, and N. M.
Bronnikova, Astron. Astrophys. {\bf 418}, 357 (2004).

\noindent 16. E.V. Khrutskaya and M. Yu. Khovritchev, Izv. Gl.
Astron. Obs. {\bf 217}, 337 (2004).

\noindent 17. J. Kovalevsky, L. Lindegren, M.A.C. Perryman, et
al., Astron. Astrophys. {\bf 323}, 620 (1997).

\noindent 18. C. Ma, E. F. Arias, T.M. Eubanks, et al., Astrophys.
J. {\bf 116}, 516 (1998).

\noindent 19. S.R. Majewski, M.F. Skrutskie, M.D. Weinberg, et
al., Astrophys. J. {\bf 599}, 1082 (2003).

\noindent 20. M. Miyamoto and Z. Zhu, Astron. J. {\bf 115}, 1483
(1998).

\noindent 21. M. Miyamoto and M. S\^oma, Astron. J. {\bf 105}, 691
(1993).

\noindent 22. M. Miyamoto, M. S\^oma, and M. Yoshizawa, Astron. J.
{\bf 105}, 2138 (1993).

\noindent 23. B. du Mont, Astron. Astrophys. {\bf 61}, 127 (1977).

\noindent 24. B. du Mont, Astron. Astrophys. {\bf 66}, 441 (1978).

\noindent 25. K.F. Ogorodnikov, Dynamics of Stellar Systems
(Fizmatgiz, Moscow, 1958; Pergamon, Oxford, 1965).

\noindent 26. R. Olling and W. Dehnen, Astrophys. J. {\bf 599},
275 (2003).

\noindent 27. I. Platais, V. Kozhurina-Platais, T.M. Girard, et
al., Astron. Astrophys. {\bf 331}, 1119 (1998).

\noindent 28. M.E. Popova and A.V. Loktin, Pis'ma Astron. Zh. {\bf
31}, 743 (2005) [Astron. Lett. {\bf 31}, 663 (2005)].

\noindent 29. M. Rapaport, J.-F. Le Campion, C. Soubiran, et al.,
Astron. Astrophys. {\bf 376}, 325 (2001).

\noindent
30. The HIPPARCOS and Tycho Catalogues, ESA SP-1200
(1997).

\noindent 31. M.V. Zabolotskikh, A.S. Rastorguev, and A.K. Dambis,
Pis'ma Astron. Zh. {\bf 28}, 516 (2002) [Astron. Lett. {\bf 28},
454 (2002)].

\noindent 32. N. Zacharias, S.E. Urban, M.I. Zacharias, et al.,
Astron. J. {\bf 127}, 3043 (2004). }

\medskip
{\it Translated by V. Astakhov}

\newpage
%%%%%%%%%%%%%%%%%%%%%%%%%%%%%%%%%%%%%%%% FIG.1:
\begin{figure*}[t]
{\begin{center}
  \includegraphics[width=140mm]{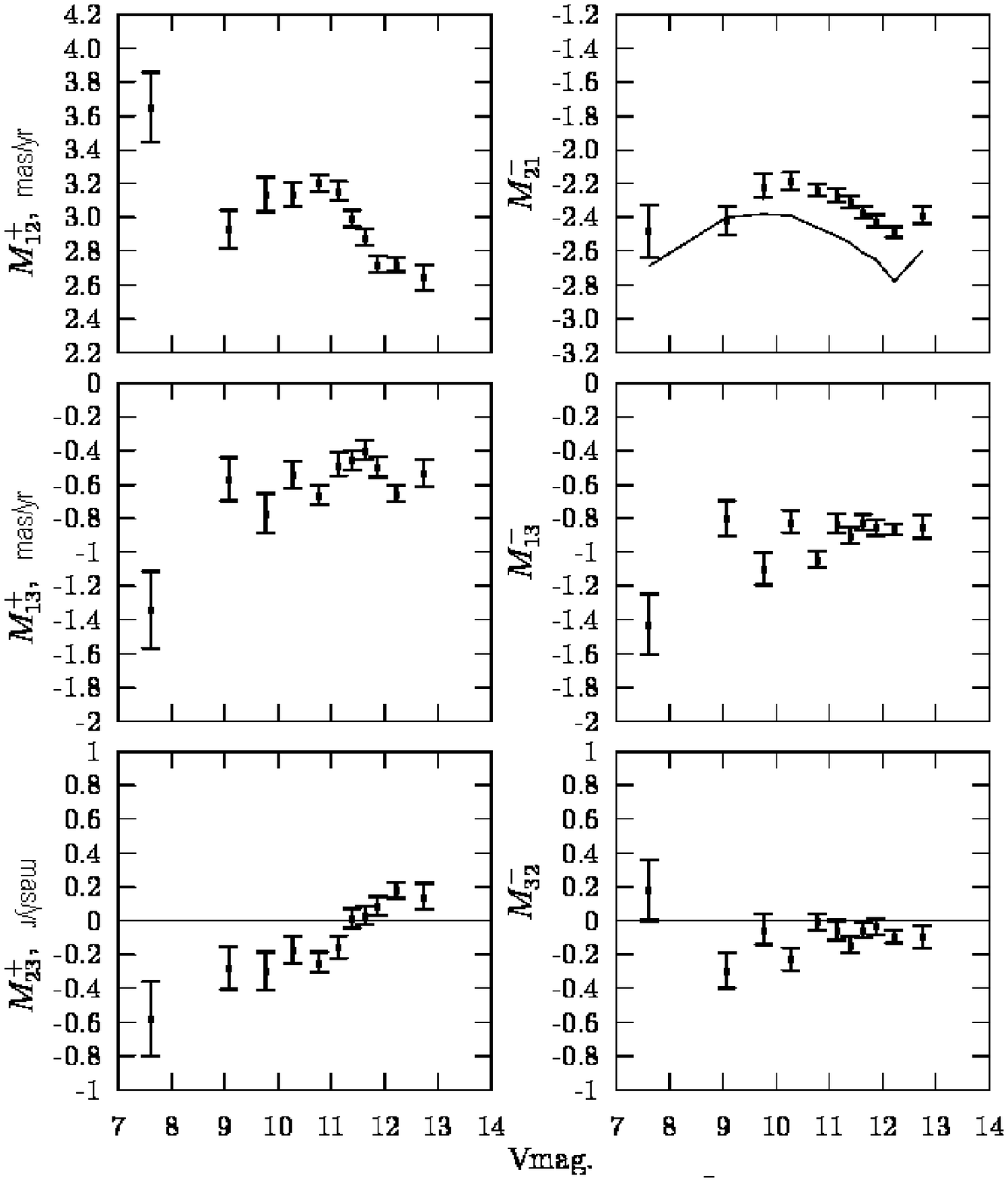}
\end{center}
\centerline {{\bf Fig. 1.} Kinematic parameters inferred from the
proper motions of Tycho-2 stars vs. magnitude. }}
\end{figure*}

\newpage
%%%%%%%%%%%%%%%%%%%%%%%%%%%%%%%%%%%%%%%% FIG.2:
\begin{figure*}[t]
{\begin{center}
  \includegraphics[width=140mm]{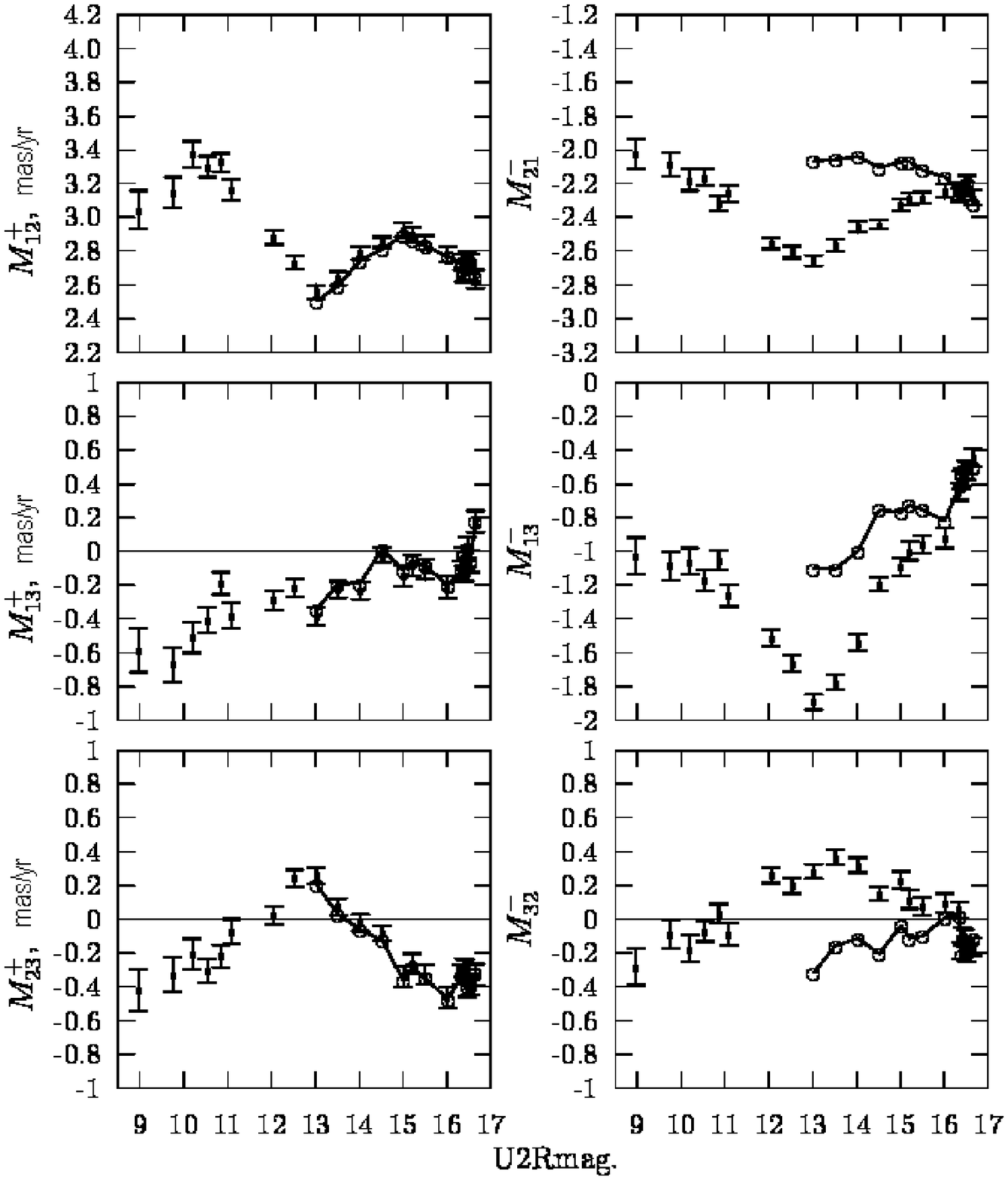}
\end{center}
\centerline {{\bf Fig. 2.} Kinematic parameters inferred from the
proper motions of UCAC2 stars vs. magnitude. }}
\end{figure*}

%%%%%%%%%%%%%%%%%%%%%%%%%%%%%%%%%%%%%%%% FIG.3: Уравнение блеска.
\newpage
\begin{figure*}[t]
{\begin{center}
  \includegraphics[width=100mm]{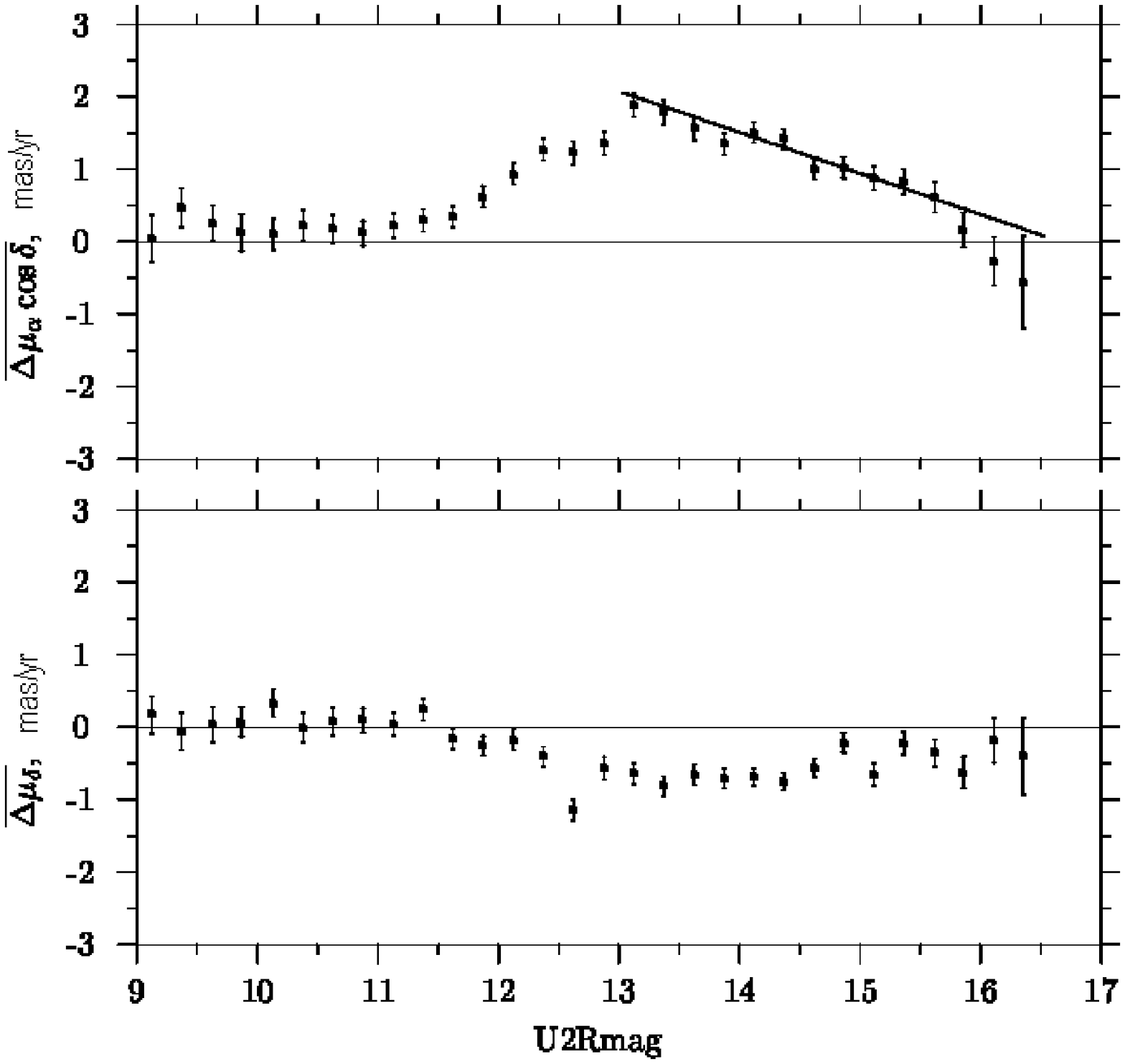}
\end{center}
\centerline {{\bf Fig. 3.} PUL3SE-UCAC2 stellar proper motion
differences vs. magnitude: (a) $\Delta\mu_\alpha\cos\delta$, (b)
$\Delta \mu_\delta$. }}
     \vskip 80mm
\end{figure*}

%%%%%%%%%%%%%%%%%%%%%%%%%%%%%%%%%%%%%%%% FIG.4: pi_W-pi_U.  HIPPARCOS.
\newpage
\begin{figure*}[t]
{\begin{center}
  \includegraphics[width=100mm]{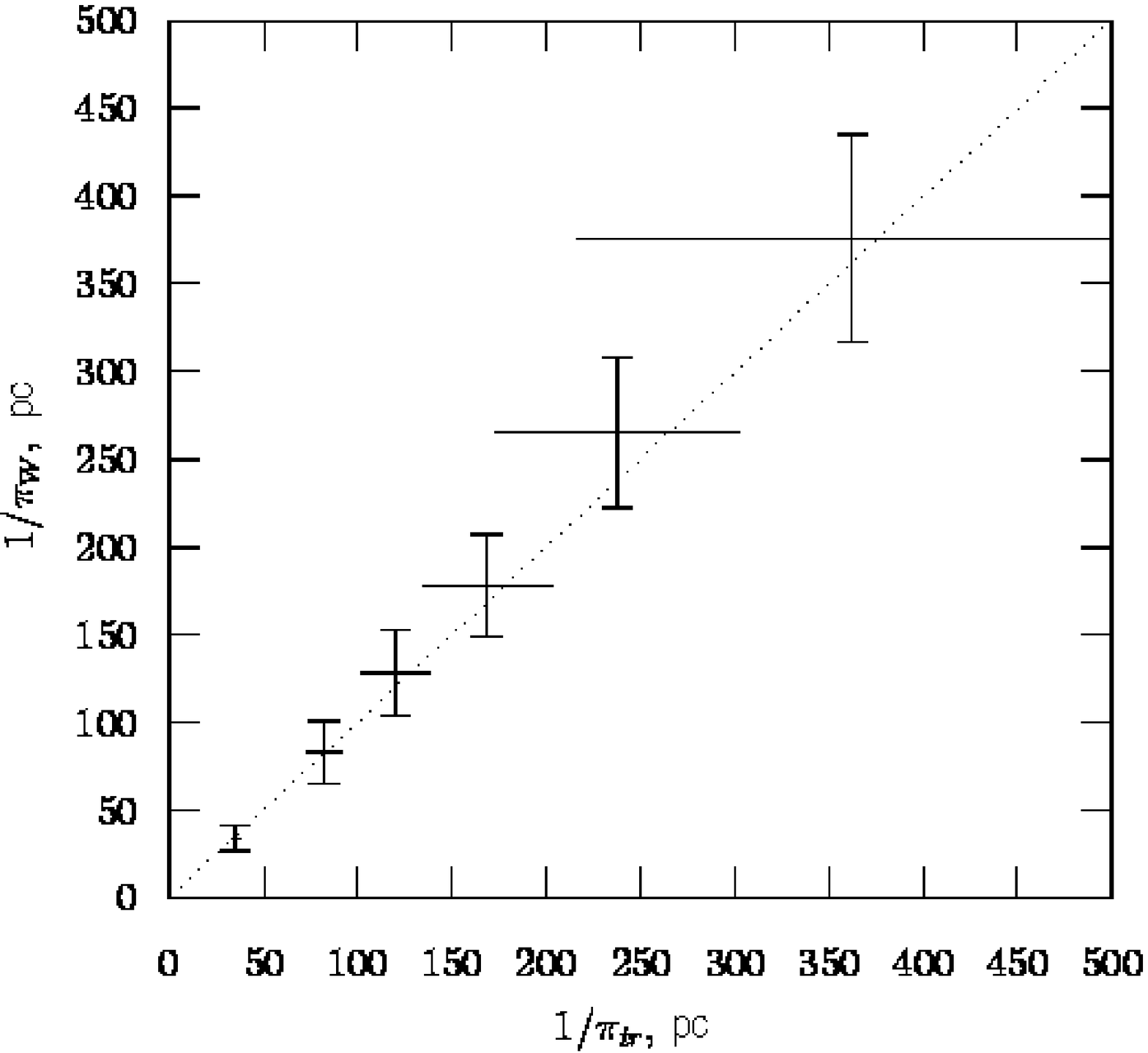}
\end{center}
{\bf Fig. 4.} Statistical distances to stars vs. trigonometric
distances calculated using Hipparcos data. The division into
groups was performed using trigonometric parallaxes. }
\end{figure*}
%%%%%%%%%%%%%%%%%%%%%%%%%%%%%%%%%%%%%%%% FIG.5:  pi_W-pi_U.  UCAC2.
%\newpage
\begin{figure*}[p]
{\begin{center}
  \includegraphics[width=100mm]{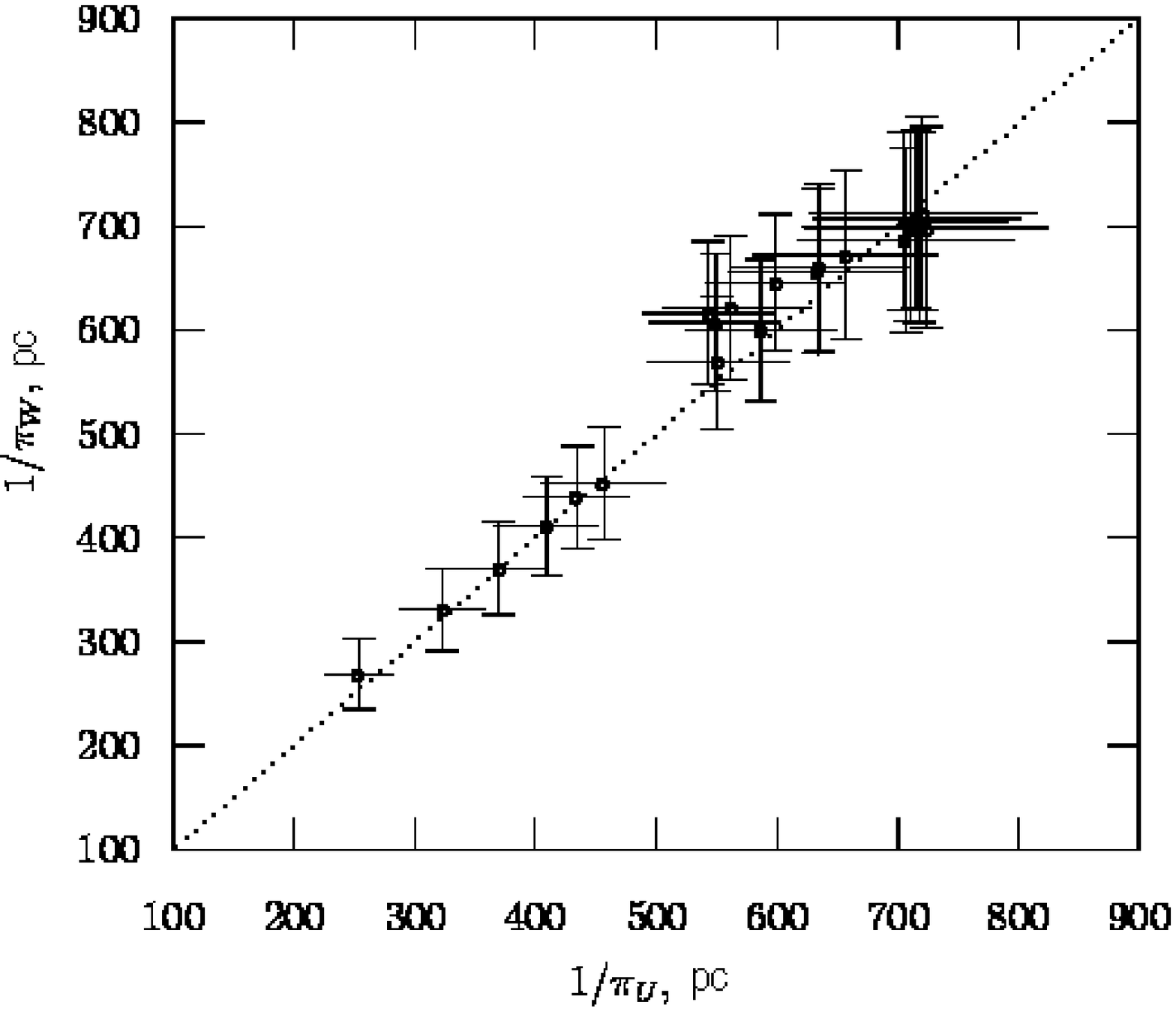}
\end{center}
{\bf Fig. 5.} Statistical distances to stars calculated from the
solar velocity component $W_\odot$ vs. distances calculated from
component $V_\odot$ using UCAC2 data. }
\end{figure*}

%%%%%%%%%%%%%%%%%%%%%%%%%%%%%%%%%%%%%%%% FIG.6:  M_y-Dist.  Tycho+UCAC2.
\newpage
\begin{figure*}[t]
{\begin{center}
  \includegraphics[width=120mm]{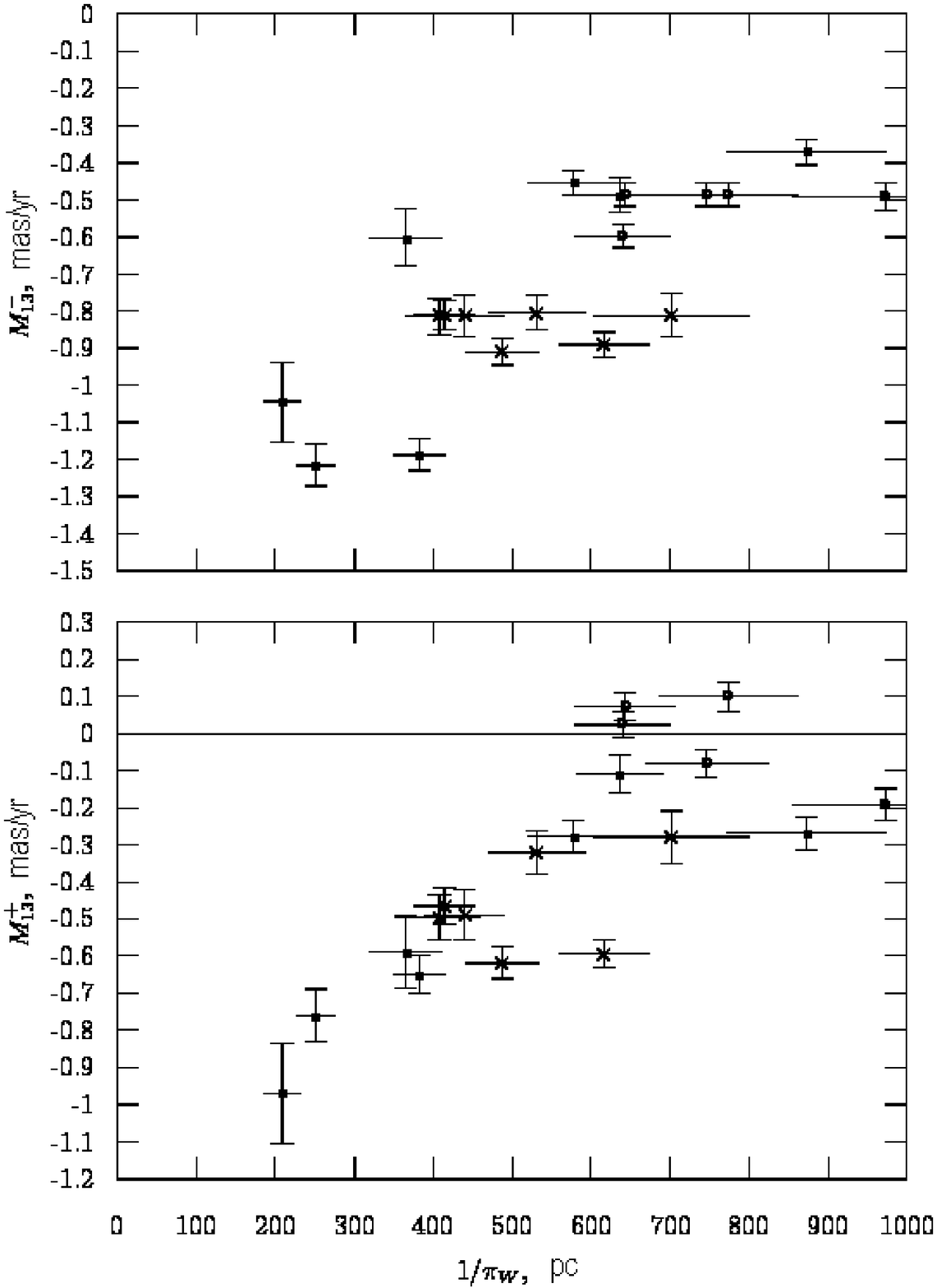}
\end{center}
{\bf Fig. 6.} Rotation, $M^-_{13}$ (a), and deformation,
$M^-_{13}$ (b), tensor components vs. distance. The dots are
bright Tycho-2 stars, the crosses are faint Tycho-2 stars, the
open circles are the faintest UCAC2 stars, and the filled circles
are bright UCAC2 stars; the horizontal bars indicate the distance
errors. }
\end{figure*}

\end{document}